# Improved crystallographic compatibility and magnetocaloric reversibility in Pt substituted $Ni_2Mn_{1.4}In_{0.6}$ magnetic shape memory Heusler alloy

K. K. Dubey[1], P. Devi[2,3], Anupam K. Singh[1] and Sanjay Singh[1*]

[1]School of Materials Science and Technology, Indian Institute of Technology (BHU), Varanasi-221005, India

[2]Max Planck Institute for Chemical Physics of Solids, Nothnitzer Str. 40, 01187 Dresden

[3]Ames Laboratory, US Department of Energy, Iowa State University, Ames, Iowa 50011, USA

**Abstract**

We present here the improved crystallographic/geometric compatibility and magnetocaloric reversibility by measurement of magnetic entropy change using different protocols in 10% Pt substituted $Ni_2Mn_{1.4}In_{0.6}$ magnetic shape memory alloy. The substitution of Pt reduces the thermal hysteresis about 50% to the $Ni_2Mn_{1.4}In_{0.6}$. The origin of the reduced thermal hysteresis is investigated by the crystallographic compatibility of the austenite and martensite phases. The calculated middle eigenvalue of the transformation matrix turned out to be 0.9982, which is very close to 1 (deviation is only 0.18%) suggests for the crystallographic compatibility between the austenite and martensite phases in $Ni_{1.9}Pt_{0.1}Mn_{1.4}In_{0.6}$. A very small thermal hysteresis and crystallographic compatibility between two phases in this alloy system indicate a stress-free transition layer (i.e. perfect habit plane) between the austenite and martensite phase, which is expected to give reversible martensite phase transition and therefore reversible magnetocaloric effect (MCE) as well. The calculated value of the isothermal entropy change ($\Delta S_{iso}$) using the magnetization curve under three different measurement protocols (i.e. isothermal, loop, and isofield measurement protocol) is found to be nearly same indicating a reversible MCE in the present alloy system. Our work provides a path to design new magnetic shape memory Heusler alloys for magnetic refrigeration and also suggest that any of the above measurement protocol can be used for the calculation of $\Delta S_{iso}$ for materials satisfying geometrical compatibility condition.

## 1. Introduction

A magnetic material heats up or cools down with the application of magnetic field in adiabatic condition and this phenomena is known as the magnetocaloric effect (MCE). MCE has created enormous interest in the solid state cooling technology [1-4] and typically presents in all the magnetic materials to some extent. The value of MCE is large near the first order magnetic phase transition (FOMT)/ first-order magnetostructural (martensite) phase transition (FOMST) where magnetization changes abruptly [5-7]. A giant MCE has been reported in various systems around their FOMT/FOMST phase transition e.g. in $Gd_5Si_2Ge_2$ [8], La-Fe-Si-based (Fe, Si)$_{13}$] [9, 10], Mn-As based [11], $Gd_5(Si_{1-x}Ge_x)_4$ [8] and magnetic shape memory Heusler alloys (e.g. $Ni_{2+x}Mn_{1-x}Z$; Z=Ga, In, Sn, and Sb) [5, 12-14]. Among these materials, magnetic shape memory Heusler alloys (MSMAs) have emerged as the suitable candidate for magnetic refrigeration applications due to their large MCE, rare earth free alloy design, no-toxic elements and their transition temperature can be easily tuned by varying composition [15, 16]. A large amount of work has been done on Ni-Mn based MSMAs aiming application as MCE material [5, 6, 12, 17-25]. The large MCE in MSMAs is mainly related to their FOMST. However, the same FOMST, which gives rise to the giant MCE, is also responsible for the irreversibility of the phase transition under the repeating magnetic field cycles and this became a major challenge in magnetic refrigeration devices [17, 21, 26-30]. This irreversibility is associated with the thermal hysteresis of FOMST due to a large stress at the transition layer between the austenite and the martensite phases. Therefore, more recently, the science community has given more focus on the minimization of the hysteresis [19, 21, 31-33]. Hysteresis is the intrinsic nature of the materials undergoing FOMT/FOMST, and the one way to reduce hysteresis is by substituting the extra atom (few %) on a particular atomic site [7, 16]. Recently, a study on the structural and magnetic properties of Pt substituted $Ni_2MnGa$ [7] is done, where it has been observed that the replacement of the Ni by Pt not only brings the martensite transition temperature close to the room temperature but also reduces the thermal hysteresis [7]. It has been proposed later that the Pt substitution might facilitate the requirement of the invariant habit plane [34], which supports shape memory behavior i.e. reversible martensite transition. This suggests that Ni-Pt-Mn-Ga are good candidate for magnetic actuators and magnetic cooling applications. Comparing the MCE of different Ni-Mn based MSMAs it turns out that $Ni_2Mn_{1.4}In_{0.6}$ is one of the best composition for the MCE application due to its large entropy and adiabatic temperature change although the irreversibility with repeating magnetic field cycle is a major challenge [6, 17, 21, 23, 30, 35-37]. Therefore, we selected this composition for our study and investigated the effect of 10% Pt substitution in $Ni_2Mn_{1.4}In_{0.6}$ MSMA. $Ni_2Mn_{1.4}In_{0.6}$ shows martensite transition near room temperature with a thermal hysteresis of ~8 K [38]. A 10% substitution of Pt in $Ni_2Mn_{1.4}In_{0.6}$ reduces thermal hysteresis about 50% and the obtained MSMA composition i.e. $Ni_{1.9}Pt_{0.1}Mn_{1.4}In_{0.6}$ exhibits a narrow thermal hysteresis ~4 K. We investigated the origin of the reduced hysteresis via calculation of the structural transformation matrix and crystallographic compatibility factor based on the lattice parameters obtained from the structural analysis of synchrotron X-ray powder diffraction (SXRPD) data in both austenite and martensite phases. The middle eigenvalue ($\lambda_2$) of the transformation matrix is found to be very close to 1 and smaller deviation from unity than the previously reported values in these alloys [21, 39, 40] suggests a low energy barrier or the stress free transition layer between austenite and martensite phases, which provides geometrical compatibility between two phases. Therefore, it is expected to show a reversible phase transition and MCE [40]. An indirect evidence of reversibility is obtained from the calculation of magnetic entropy change using different measurement protocols as suggested in the literature [41]. The similar value of magnetic entropy change further indicates the reversibility of MCE.

2. Experimental

A polycrystalline ingot of $Ni_{1.9}Pt_{0.1}Mn_{1.4}In_{0.6}$ was prepared by arc melting in argon atomosphere [12, 23, 42]. The sample was melted four times for homogeneity and subsequently annealed in an evacuated sealed quartz ampoule at 900 $^0C$ for 24 h and then quenched in the ice water mixture. For the characterization, a part of the bulk sample was crushed into powder and then annealed under vacuum to remove residual stress-induced effects generated after crusing [43-45]. Magnetization measurement was performed by using magnetic property measurement system (Quantum Design) at a low magnetic field of 0.05 T in the temperature range 100 K to 375 K to determine the transition temperatures, range of transition and thermal hysteresis at the phase transition. To investigate the crystal structure and to determine the crystallographic compatibility condition, SXRPD data were recorded in both austenite and martensite phases using P02 beamline at Petra III, Hamburg, Germany by using a wavelength of 0.20712 Å. Further to calculate $\Delta S_{iso}$, magnetization data were recorded using magnetic property measurement system (Quantum Design) in three measurement protocols named as isothermal, loop, and isofield protocol [46-50]. In isofield measurement protocol magnetization (*M*) vs temperature (*T*) data in the range of 150 K to 400 K were recorded during field cooled cooling (FCC) and field cooled warming (FCW) at different magnetic field values of 1 T, 3 T, 4 T, and 7 T. Magnetization vs field (*H*) data were recorded in isothermal and loop measurement protocols from the temperature range 210 K to 321 K at the step of 3 K up to 7 T applied field. For the simplicity of figure we show here *M (H)* data around martensite phase transition region (231 K to 273 K) at the step of 3 K up to 7 T applied field.

## 3. Results and discusssion

The magnetization ($M\ (T)$) curves for the parent compound ($Ni_2Mn_{1.4}In_{0.6}$) and Pt doped compound, i.e. $Ni_{1.9}Pt_{0.1}Mn_{1.4}In_{0.6}$ under a low applied magnetic field of 0.05 T are shown in figure 1(a) and figure 1(b), respectively. The $M\ (T)$ curve for the parent compound (figure 1(a)) is obtained from Ref [38]. Both the alloys undergo magneto-structural (martensite) phase transition upon cooling (FCC curve) and reverse martensite transition during heating (FCW curve). The hysteresis observed during heating and cooling is the signature of the first-order phase transition. The characteristic temperatures related to the martensite phase transformations, i.e. martensite start ($M_s$), martensite finish ($M_f$), austenite start ($A_s$), and austenite finish temperature ($A_f$) for the $Ni_2Mn_{1.4}In_{0.6}$ are 296 K, 221 K, 232 K, and 301 K, respectively (figure 1(a)). For $Ni_{1.9}Pt_{0.1}Mn_{1.4}In_{0.6}$ these temperatures, i.e. $M_s$, $M_f$, $A_s$, and $A_f$ are 281 K, 238 K, 243 K, and 284 K, respectively (figure 1(b)). The ferromagnetic transition temperature ($T_C$) for $Ni_2Mn_{1.4}In_{0.6}$ and $Ni_{1.9}Pt_{0.1}Mn_{1.4}In_{0.6}$ are ~ 314 K and ~ 304 K, respectively. The width of thermal hysteresis of the martensite transition is calculated from the martensite and austenite characteristic transition temperatures ($M_s$, $M_f$ and $A_s$, $A_f$) using the formula $\frac{[(A_f+A_s) \sim (M_f+M_s)]}{2}$. The thermal hysteresis for the parent compound ($Ni_2Mn_{1.4}In_{0.6}$) is ~ 8 K, while hysteresis reduces to ~ 4 K for $Ni_{1.9}Pt_{0.1}Mn_{1.4}In_{0.6}$ (figure 1(b)). This value of thermal hysteresis is smaller in comparison to the other reported MSMAs [21, 39, 41]. Recently a reversible MCE is reported in $Ni_{2.2}MnGa$ MSMA, where thermal hysteresis around martensite transition is observed about 5 K [41]. It has been suggested in the literature that the thermal hysteresis observed around the martensite transition is a direct consequence of the structural compatibility between austenite and martensite phases, which is known as the crystallographic compatibility criteria [21, 39, 41, 51-56]. This crystallographic compatibility between austenite and martensite phases depends on the crystal structure/unit cell parameters of the austenite and martensite phases [55-58]. In order to investigate the crystal structure of the austenite and martensite phases and to get the actual unit cell parameters, temperature dependent SXRPD data were collected in both the austenite phase (350 K) and martensite phase (110 K). Le Bail refinements of SXRPD data of both the phases were done (figure 2 and figure 3) using Fullprof software package [59]. The austenite phase was refined by cubic L2$_1$ structure (space group F*m*-3*m*) as expected in these alloys [23]. The refined lattice parameter turned out to be 6.0149 Å. Figure 2 shows that the observed and calculated peak profile are well matched, confirming the cubic structure of the austenite phase. The diffraction pattern of the martensite phase (figure 3) has large number of peaks indicating a modulated martensite phase as reported in the literature for Ni-Mn based MSMAs [7, 42, 60]. It has been suggested that the low temperature structure of parent compound ($Ni_2Mn_{1.4}In_{0.6}$) is 3*M* modulated monoclinic with space group I*2/m* [38]. A detailed investigation of martensite structure after 10% Pt doping in parent compound i.e. $Ni_{1.9}Pt_{0.1}Mn_{1.4}In_{0.6}$ was done using Le Bail refinements. Firstly, the structure of parent compound (i.e. 3*M* modulated monoclinic) was considered during the refinement but a clear mismatch between observed and calculated profiles was noticed as shown in figure 3(a). Some of the peaks were not indexed and they are indicated by arrows in the inset of figure 3(a). In the next step to identify these peaks, a larger unit cell i.e. 5*M* modulated unit cell was considered but this structure was also unable to identify all the Bragg peaks (shown in the inset of figure 3(b)). After trying many crystal structure combinations (not shown here), we found that all the Bragg reflections of the martensite phase at 110 K can be indexed by a monoclinic unit cell (space group P*2/m*) and lattice parameter *a*=4.4172Å, *b*=5.6102Å, *c*=13.0350Å, and *β*=93.361$^0$ (figure 3(c)).

Here, $c\sim3*a$ indicating a 3M modulated monoclinic structure of the martensite phase [42, 60-62] of $Ni_{1.9}Pt_{0.1}Mn_{1.4}In_{0.6}$.

Having obtained the crystal structure and lattice parameters of the austenite and martensite phases of $Ni_{1.9}Pt_{0.1}Mn_{1.4}In_{0.6}$, now we turn the discussion towards the investigation of crystallographic compatibility criteria proposed for the reversible martensite transformation. It has been reported that the lattice parameter of the high symmetry austenite phase and low symmetry martensite phase are related by a transformation matrix U (equation (1)) [54, 55, 58, 63, 64], which is a 3x3 homogenous matrix defined as deformation matrix [21, 58, 65]. In the case of the complete reversibility of the martensite transformation, the $\lambda_2$ of the U should approach 1. The components of the transformation matrix depend entirely on the lattice parameters of the austenite (high temperature) and martensite (low temperature) phases, as described below [40, 41, 52, 53, 63, 65, 66].

$$U = \begin{bmatrix} \tau & \sigma & 0 \\ \sigma & \rho & 0 \\ 0 & 0 & \delta \end{bmatrix} \quad (1)$$

The components of the transformation matrix ($\tau$, $\rho$, $\sigma$, and $\delta$) are expressed as follows:

$$\tau = \frac{\alpha^2 + \gamma^2 + 2\alpha\gamma(\sin\beta - \cos\beta)}{2\sqrt{\alpha^2 + \gamma^2 + 2\alpha\gamma(\sin\beta)}} \quad (2)$$

$$\rho = \frac{\alpha^2 + \gamma^2 + 2\alpha\gamma(\sin\beta + \cos\beta)}{2\sqrt{\alpha^2 + \gamma^2 + 2\alpha\gamma(\sin\beta)}} \quad (3)$$

$$\sigma = \frac{\alpha^2 - \gamma^2}{2\sqrt{\alpha^2 + \gamma^2 + 2\alpha\gamma(\sin\beta)}} \quad (4)$$

$$\delta = \frac{b}{a_0} \quad (5)$$

where, $\alpha = \sqrt{2}\frac{a}{a_0}$, $\gamma = \sqrt{2}\frac{c}{Na_0}$, $a_0$ is the lattice parameter of the austenite (cubic) unit cell and $a$, $b$, $c$, and $\beta$ denote the lattice parameter of the martensite (monoclinic) unit cell. $N$ is the degree of modulation (in the present case $N=3$). Using the lattice parameters obtained from the Le Bail refinements of the SXRPD data we calculated the components of the U for $Ni_{1.9}Pt_{0.1}Mn_{1.4}In_{0.6}$ and which turned out as:

$$U = \begin{bmatrix} 1.0598 & 0.0085 & 0 \\ 0.0085 & 0.9994 & 0 \\ 0 & 0 & 0.9327 \end{bmatrix}$$

The eigenvalues of the U are then calculated by simple mathematical calculations for a matrix. The eigen values turned out to be $\lambda_1=0.9327$, $\lambda_2=0.9982$, and $\lambda_3=1.0610$. Thus, the middle eigen value ($\lambda_2$) is 0.9982, which is nearly equal to 1 and its deviation from unity is only 0.0018 (i.e. 0.18%), which is smaller than previously reported value [21, 39, 40, 51]. Interestingly, the value of $\lambda_2$ is smaller than that of the parent compound $Ni_2Mn_{1.4}In_{0.6}$ as well for which the value of $\lambda_2$ is 1.0042 (calculated using lattice parameters reported in Ref [38]) and therefore its deviation from unity is 0.0042 (i.e. 0.42%). Thus the analysis of SXRPD data and U together explain the origin of reduced hysteresis after 10% Pt substitution in $Ni_2Mn_{1.4}In_{0.6}$. There are some recent reports on systems satisfying crystallographic compatibility condition (i.e. the value of $\lambda_2$ is very close to 1) and gives rise to a lower value of thermal hysteresis, which results into a reversible MCE [33, 67, 68]. These results clearly show that $Ni_{1.9}Pt_{0.1}Mn_{1.4}In_{0.6}$ has low thermal hysteresis and also crystallographic compatibility criteria has been improved. Therefore, in this alloys system, a field induced reversible martensite phase transition and hence, the reversible MCE is expected [33]. In general, there are two methods to estimate MCE. The first method is the "direct measurement method" where the adiabatic temperature change is measured directly under external applied magnetic field cycles [17, 21, 40, 41]. The second method is known as the "indirect measurement method" where $\Delta S_{iso}$ is measured using the magnetic hysteresis loop. For the indirect measurement of MCE different measurement protocols have been suggested for the calculation of the $\Delta S_{iso}$ especially for the FOMT/FOMST [20, 27, 48, 50, 69-75]. These measurement protocols are not only used to comment on the correct value of $\Delta S_{iso}$ but also useful to comment on the reversibility/irreversibility of MCE [33, 49]. If these indirect measurement protocols provide a similar value of $\Delta S_{iso}$ then a reversible MCE is expected [46]. Therefore, now we proceed for the calculation of the $\Delta S_{iso}$ using magnetization measurement under three different protocols reported in the literature and named as isothermal, loop, and isofield protocol as discussed in the experimental section [46-48, 70, 76]. Following isothermal protocol, isothermal *M (H)* curves are shown around the phase transition region between 231 K to 273 K with 3 K intervals (figure 4(a)). For loop measurement protocol the sample was heated up to 400 K to ensure the full austenite phase followed by cooling without application of the magnetic field down to 200 K to ensure the complete martensite phase transition and then subsequently sample was heated up to the desired measurement temperature where the *M (H)* data were recorded (figure 4(b)) [41, 48, 70, 76]. Using these *M (H)* curves, $\Delta S_{iso}$ is calculated using Maxwell's equation.

$$\Delta S_{iso} = \mu_0 \int_0^H \left(\frac{\partial M}{\partial T}\right)_H dH \qquad (6)$$

For isofield measurement protocol, *M (T)* during heating and cooling was recorded at different representative fields (1 T, 3 T, 4 T, and 7 T) as shown in the figure 4(c). For the $\Delta S_{iso}$ calculation, *M (H)* data were extracted at different temperatures around the phase transition regions from these isofield curves manually. To take *M (H)* data from the isofield curve, first the temperature was fixed, and the value of *M* and corresponding *H* from all isofield curves were noted down. This procedure provided the *M (H)* curve at a fixed temperature. Similarly, *M (H)* data is extracted for all other temperatures, which provided a series of the *M (H)* curves at different temperatures (figure 4(d)). Then using Maxwell's relation (equation (6)) $\Delta S_{iso}$ was calculated for different field values similar to the other two (isothermal and loop) measurement protocols. The calculated value of $\Delta S_{iso}$ as a function of temperature is compared for all three measurement protocols at field values of 1 T, 3 T, 4 T, and 7 T (figure 5). It is interesting to note that the values of the $\Delta S_{iso}$ at different fields are unambiguously the same for all three measurement protocols suggesting a reversible MCE in $Ni_{1.9}Pt_{0.1}Mn_{1.4}In_{0.6}$ under the magnetic field cycles [41].

To summarize, we have shown the improved crystallographic compatibility and reversible magnetocaloric using different measurement protocols in 10% Pt substituted $Ni_2Mn_{1.4}In_{0.6}$ magnetic shape memory Heusler alloy. The 10% Pt in $Ni_2Mn_{1.4}In_{0.6}$ reduces thermal hysteresis from ~8 K to ~4 K. The origin of the reduced thermal hysteresis is found to be related with the reduced middle eigenvalue $\lambda_2$ of the phase transformation matrix U. The calculated middle eigenvalue of the U is very close to 1 (0.9982) suggesting crystallographic compatibility between austenite and martensite phases. Moreover, nearly the same $\Delta S_{iso}$ values calculated under three different measurement protocols (isothermal, loop, and isofield protocol) indicate a reversible MCE in $Ni_{1.9}Pt_{0.1}Mn_{1.4}In_{0.6}$. However, a direct adiabatic temperature change measurement (magnetocaloric) under magnetic field cycle will be further useful to put this alloy system forward for magnetic cooling device applications. Our study shows that the energy barrier or the stress of the transition layer between austenite and martensite phases in magnetic shape memory Heusler alloys can be reduced via designing suitable composition, which provides crystallographic compatibility and the invariant habit plane between two crystallographic phases and hence a reversible martensite phase transition and magnetocaloric effect.

*ssingh.mst@iitbhu.ac.in

**Acknowledgments-**

We thank C.F. for encouragement and L.C. for useful discussion. SS thanks Science and Engineering Research Board of India for financial support through the award of Ramanujan Fellowship (grant no: SB/S2IRJN-015/2017), Early Career Research Award (grant no:

ECR/2017/003186) and UGC-DAE CSR, Indore for financial support through "CRS" Scheme. KKD thanks DST for providing fellowship through DST-INSPIRE scheme.

**Figures**

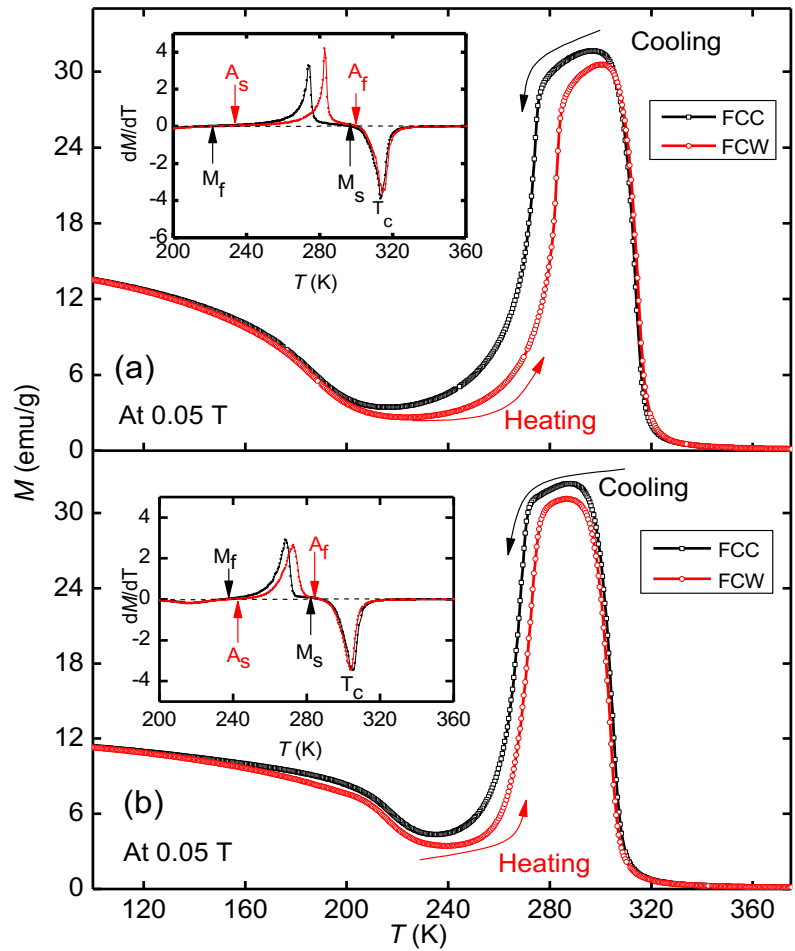

Figure 1. Temperature dependent magnetization curve of (a) $Ni_2Mn_{1.4}In_{0.6}$ and (b) $Ni_{1.9}Pt_{0.1}Mn_{1.4}In_{0.6}$ at $H= 0.05$ T. Insets show differential (d$M$/d$T$) curves for better presentation of transition and hysteresis.

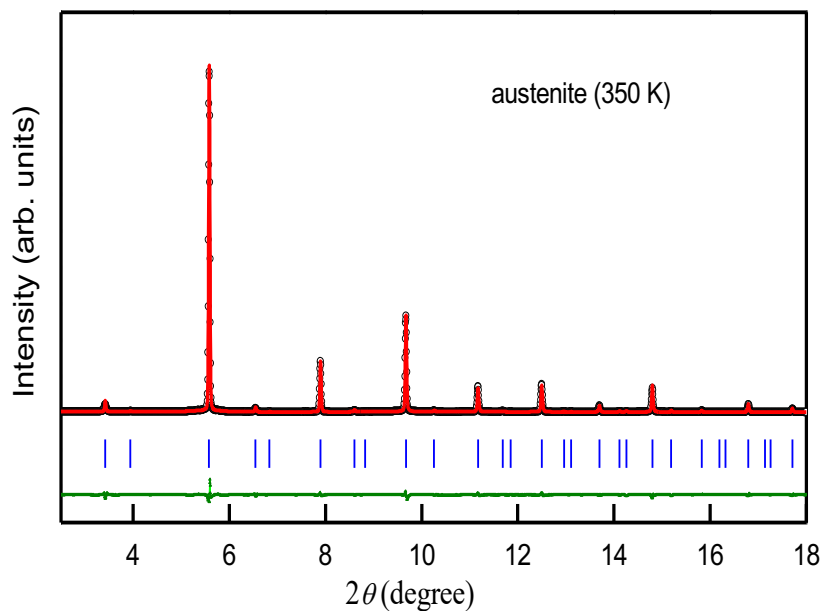

Figure 2. Le Bail refinement of SXRPD data of $Ni_{1.9}Pt_{0.1}Mn_{1.4}In_{0.6}$ in austenite phase (350 K). The experimental peak profile, calculated peak profile and the difference are shown by black circle, red and green solid lines, respectively. The blue lines represent the Bragg's peak positions.

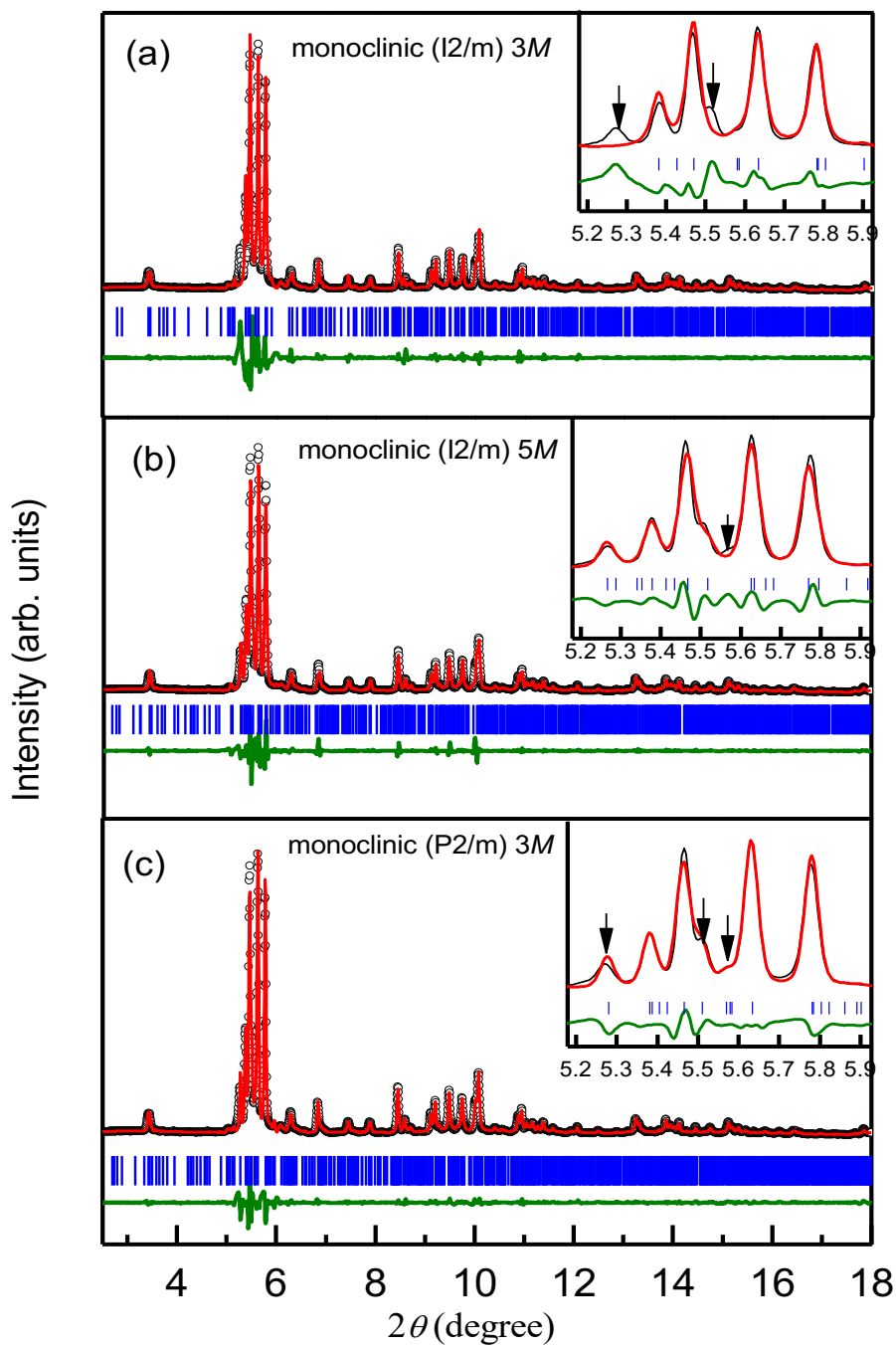

Figure 3. Le Bail refinements of SXRPD data of $Ni_{1.9}Pt_{0.1}Mn_{1.4}In_{0.6}$ in the martensite phase at 110 K using (a) Monoclinic *3M* modulated structure (space group: I*2/m*) (b) Monoclinic *5M* modulated structure (space group: I*2/m*) (c) Monoclinic *3M* modulated structure (space group: P*2/m*). The experimental peak profile, calculated peak profile and difference are shown by black circle, red and green solid line, respectively. The blue lines represent the Bragg's peak positions. In the insets of (a) and (b), arrows indicate the unindexed peak. In the inset of (c), arrows show the indexed peak which were unindexed in the inset of (a) and (b).

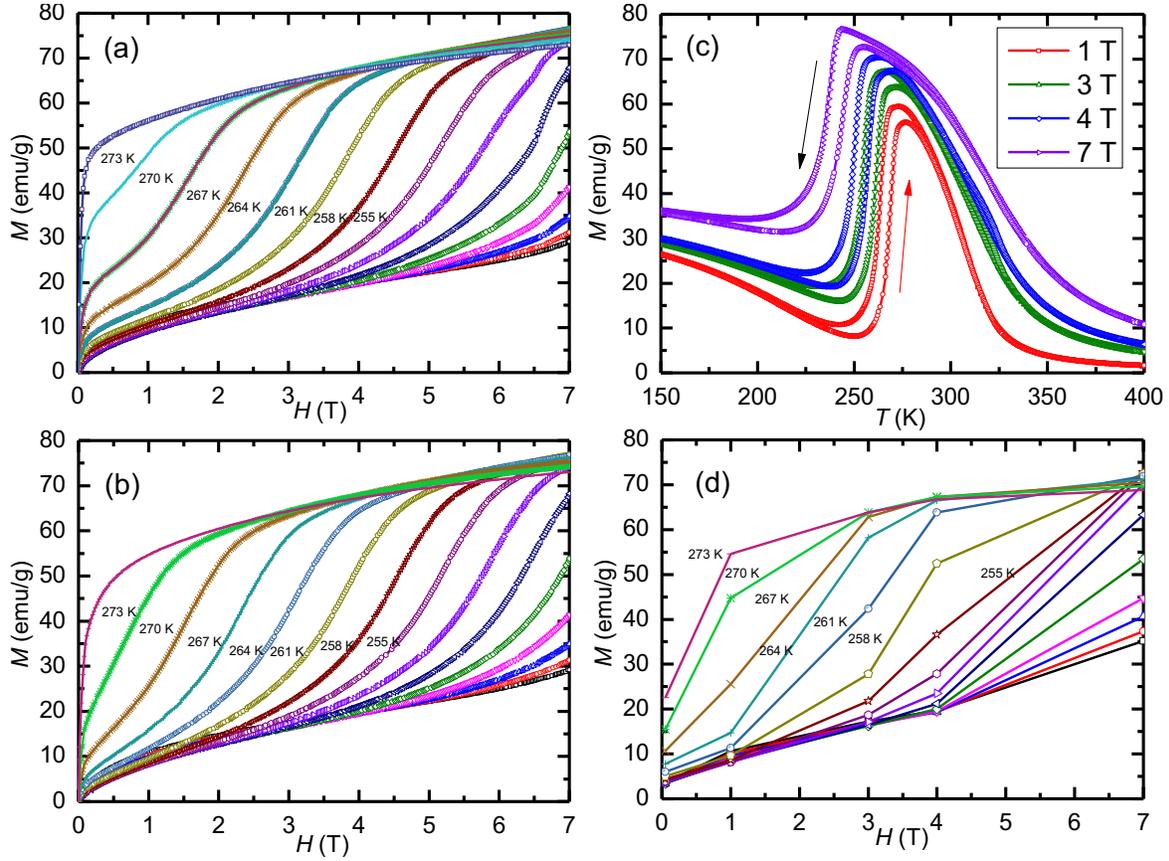

Figure 4. Magnetization measurement results for the $Ni_{1.9}Pt_{0.1}Mn_{1.4}In_{0.6}$ (a) magnetic isotherms obtained from isothermal measurent protocol. (b) magnetic isotherms obtained from the loop measurent protocol. (c) isofield (M(T)) curves of $Ni_{1.9}Pt_{0.1}Mn_{1.4}In_{0.6}$ at different field values. (d) magnetic isotherms extracted from isofield curves shown in (c).

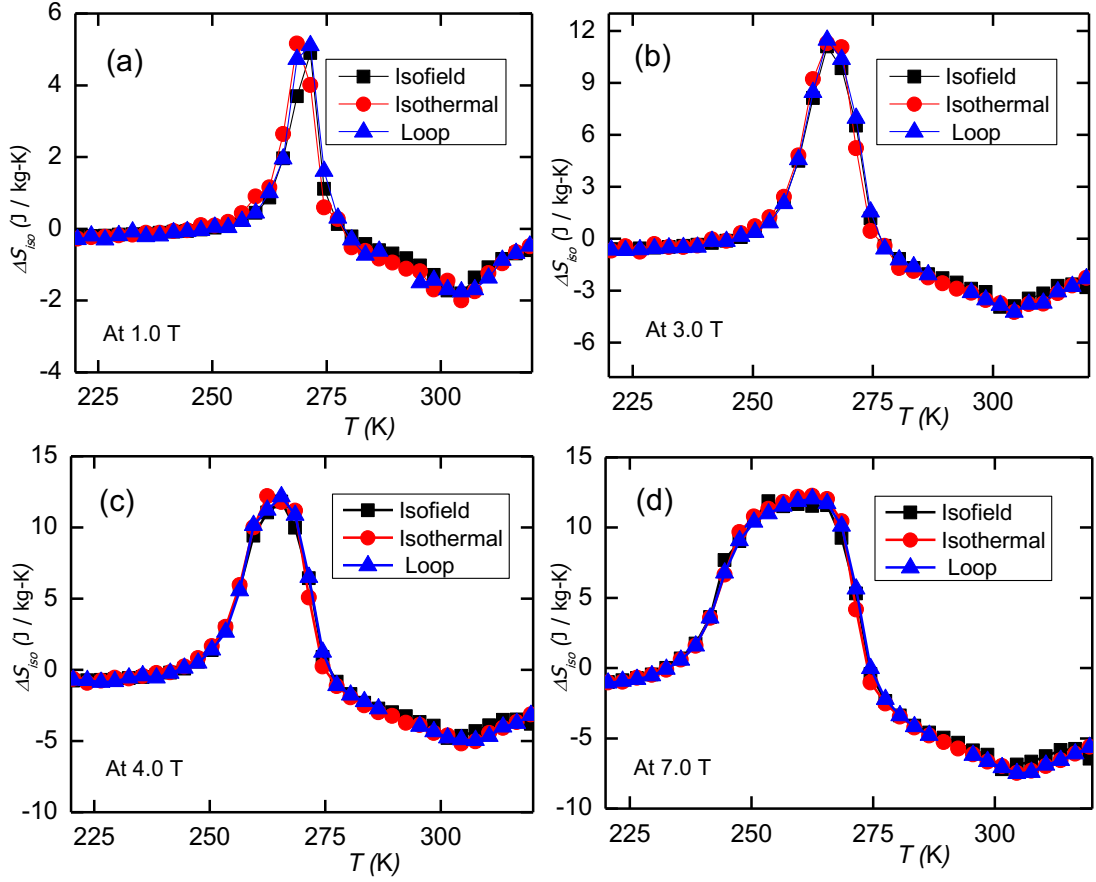

Figure 5. Comparison of isothermal entropy change $\Delta S_{iso}$ as a function of temperature for the isofield, isothermal, and loop measurement protocols for $Ni_{1.9}Pt_{0.1}Mn_{1.4}In_{0.6}$ at different applied field values (a) 1 T (b) 3 T (c) 4 T and (d) 7 T.